\documentclass[twocolumn,superscriptaddress,showpacs]{revtex4}
\usepackage{amssymb}

\usepackage{graphicx}
\usepackage{dcolumn}
\usepackage{bm}
\usepackage{stmaryrd}

\input{tcilatex}

\begin{document}

\title{Asymmetric noise probed with a Josephson junction}
\author{Q. Le Masne}
\author{H. Pothier}
\email[Corresponding author~: ]{hugues.pothier@cea.fr}
\affiliation{Quantronics group, Service de Physique de l'Etat Condens\'{e} (CNRS URA
2464), CEA-Saclay, 91191 Gif-sur-Yvette, France}
\author{Norman O. Birge}
\affiliation{Department of Physics and Astronomy, Michigan State University, East
Lansing, MI 48824, USA}
\author{C. Urbina}
\author{D. Esteve}
\affiliation{Quantronics group, Service de Physique de l'Etat Condens\'{e} (CNRS URA
2464), CEA-Saclay, 91191 Gif-sur-Yvette, France}
\date{\today }

\begin{abstract}
Fluctuations of the current through a tunnel junction are measured using a
Josephson junction. The current noise adds to the bias current of the
Josephson junction and affects its switching out of the supercurrent branch.
The experiment is carried out in a regime where switching is determined by
thermal activation. The variance of the noise results in an elevated
effective temperature, whereas the third cumulant, related to its asymmetric
character, leads to a difference in the switching rates observed for
opposite signs of the current through the tunnel junction. Measurements are
compared quantitatively with recent theoretical predictions.
\end{abstract}

\pacs{72.70.+m, 85.25.Cp, 73.23.-b, 74.50.+r}
\maketitle

The current through voltage-biased electrical conductors exhibits
fluctuations, which, in contrast to equilibrium Johnson-Nyquist noise, are
not symmetric with respect to the average current. This translates into
finite odd cumulants in the distribution of the number of electrons
transfered through the conductor in a given time. Whereas the full counting
statistics of this number can be calculated for arbitrary conductors \cite%
{Levitov,Nazarov}, up to now high order cumulants have been measured in very
few experiments. The third cumulant has been successfully accessed by signal
processing the time-dependent current \cite{Reulet,Bomse,Gershon}, but with
set-ups that are restricted either to low impedance samples, which leads to
large environmental effects \cite{Reulet}, or to low frequencies \cite%
{Bomse,Gershon}. Another strategy, put forward by Tobiska and Nazarov \cite%
{Tobiska}, consists in using a Josephson junction (JJ) as a large bandwidth
on-chip noise detector \cite{Timofeev,Peltonen,Huard}. It has a high
intrinsic sensitivity, and can be coupled to noise sources over a large
range of impedances. The detection principle relies on the exponential
current sensitivity of the switching of a JJ from a metastable zero-voltage
branch to a dissipative one. When biased at a current $I_{J}$ slightly below
its critical current $I_{0},$ the rate of switching is therefore very
sensitive to noise in the current. The first detection of asymmetric noise
with a JJ was reported in Refs.~\cite{Timofeev, Peltonen}. However, the
detector JJ, which was placed in an inductive environment, had a very large
plasma frequency, and the dynamics of the junction changed regime as the
noise intensity increased, from macroscopic quantum tunneling (MQT) to
retrapping \cite{Melnikov} through thermal activation. The measured
asymmetry in the escape rates could only be compared to an adiabatic model %
\cite{Comparison}, using empirical parameters. For a detector to be of
practical use it must have a well characterized and a simple enough
response, so that information on the noise can be reliably extracted. As
quantitative theories have been developed for a JJ in the thermal regime
placed in a resistive environment \cite{Ankerhold,Sukhorukov,Grabert}, we
designed an experiment in this framework, allowing for a detailed,
quantitative comparison with theory.

The principle of our experiment is to add the current noise from a noise
source to the DC bias current of a JJ (see Fig.\thinspace 1). The dynamics
of a JJ placed in a resistive environment are described by the celebrated
RCSJ model \cite{Barone}, with the voltage related to the average velocity
of a fictitious particle placed in a tilted washboard potential.\ The tilt
of the potential is determined by the reduced parameter $s=I_{J}/I_{0}$. At $%
s<1$ the potential presents local minima where the particle can be trapped.\
The voltage is then zero: this corresponds to the supercurrent branch. The
frequency of small oscillations is called the plasma frequency $\omega _{p}.$
Johnson-Nyquist current noise related to the finite temperature $T$ of the
environment of the junction is modeled as a fluctuating force on the
particle, which triggers escape from the local minimum (``switching''). When 
$k_{B}T>\hbar \omega _{p}/2\pi ,$ the switching rate $\Gamma $ is given by
Kramer's formula \cite{bookAnkerhold} $\Gamma =A\exp -B_{2}(T)$ with $%
A\simeq \omega _{p}/2\pi $ for moderate quality factors $Q$, $\omega
_{p}=\omega _{p0}\left( 1-s^{2}\right) ^{1/4}$ the plasma frequency in the
tilted potential$,$ $\omega _{p0}=\sqrt{I_{0}/\varphi _{0}C_{J}}$ the bare
plasma frequency determined by the capacitance $C_{J}$ and critical current $%
I_{0}$ of the junction, with $\varphi _{0}=\hbar /2e$, and%
\begin{equation}
B_{2}(T)=\left( 4\sqrt{2}I_{0}\varphi _{0}/3k_{B}T\right) (1-\left| s\right|
)^{3/2}.  \label{B2}
\end{equation}%
Recently, this result was extended to the situation in which an additional
delta-correlated noise $\delta I_{N}(t)$, characterized by a finite third
cumulant $S_{3}$ defined by $\left\langle \delta I_{N}(t)\delta
I_{N}(t^{\prime })\delta I_{N}(t^{\prime \prime })\right\rangle =S_{3}\delta
\left( t^{\prime }-t\right) \delta \left( t^{\prime \prime }-t^{\prime
}\right) $ and a second cumulant $\left\langle \delta I_{N}(t)\delta
I_{N}(t^{\prime })\right\rangle =S_{2}\delta \left( t^{\prime }-t\right) $
adds to current through the JJ \cite{Ankerhold,Sukhorukov,Grabert}. The
effect of higher order cumulants is assumed to be weak. The corresponding
fluctuating force leads to a modification of the rate: $\Gamma =A\exp
-\left( B_{2}(T_{\mathsf{eff}})+B_{3}\right) .$ The second cumulant yields
an increased effective temperature $T_{\mathsf{eff}}$ given by 
\begin{equation}
2k_{B}T_{\mathsf{eff}}/R_{\parallel }=2k_{B}T/R+S_{2}.  \label{Teff}
\end{equation}%
Here, $R$ is the parallel combination of all the resistances which produce
Johnson-Nyquist noise across the junction.\ The resistance $R_{\parallel }$
characterizes the friction acting on the fictitious particle and is, in a
simple model, given by the total resistance across the junction, including
both $R$ and the resistance $R_{N}$ of the noise source. This expression
indicates that the second cumulant of noise from the noise source $S_{2}$
simply adds to the Johnson-Nyquist noise of the rest of the circuit. The
third cumulant gives rise to the additional term 
\begin{equation}
B_{3}=-S_{3}\left( \varphi _{0}/k_{B}T_{\mathsf{eff}}\right) ^{3}\omega
_{p0}^{2}\ j(s)  \label{B3}
\end{equation}%
with $j(s)$ a function of the tilt that depends on the quality factor \cite%
{Grabert}. When reversing the sign of the average current $I_{N}$ through
the noise source, $S_{2}$ remains unchanged whereas $S_{3}$ changes sign.
Therefore, the departure from $1$ of the rate ratio 
\begin{equation}
R_{\Gamma }=\Gamma \left( +I_{N}\right) /\Gamma \left( -I_{N}\right) =\exp
\left( 2\left| B_{3}\right| \right)   \label{Rg}
\end{equation}%
is a measure of non-symmetric noise $\left( S_{3}\neq 0\right) .$%
\begin{figure}[tbp]
\includegraphics[angle=-90,width=3.4in]{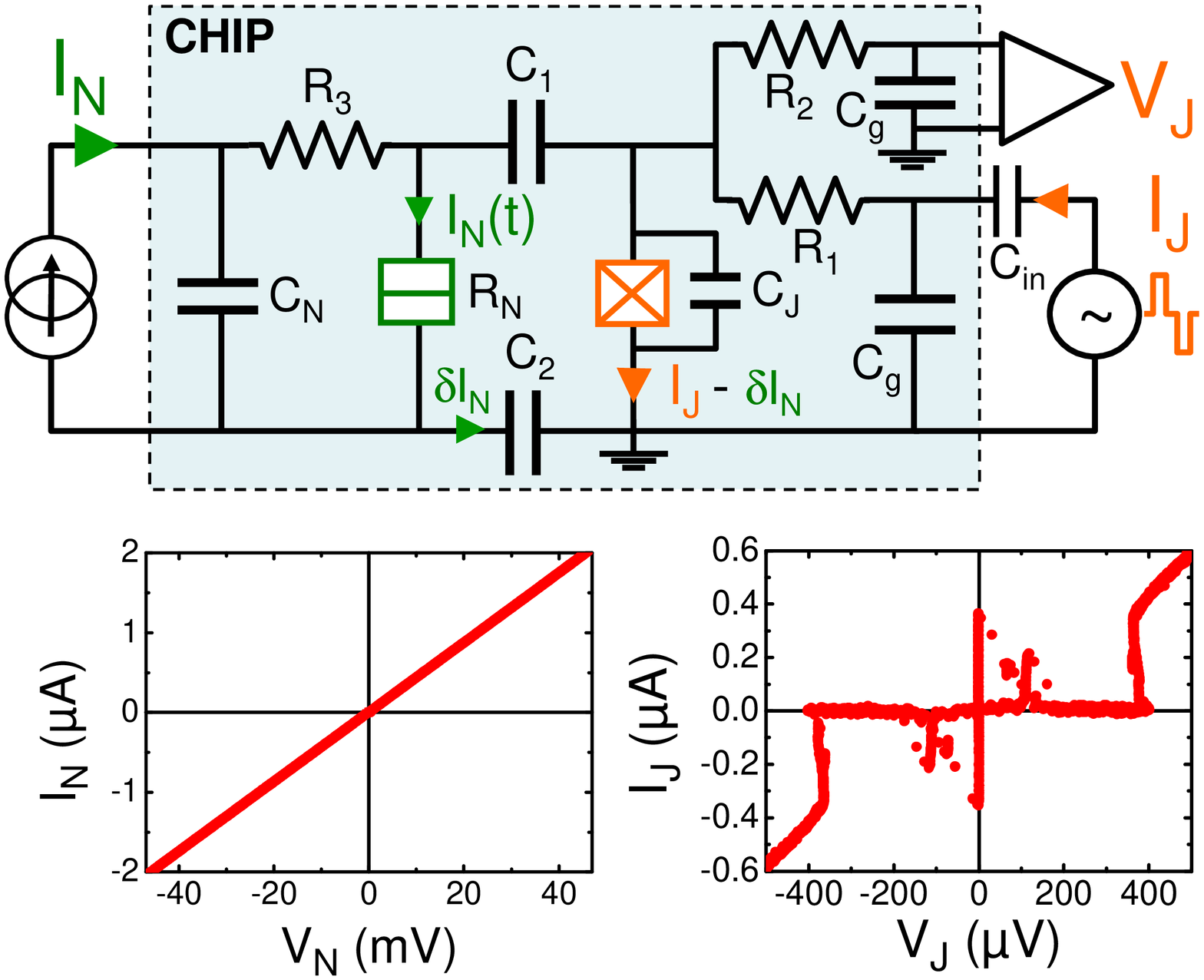}
\caption{(color online) Top: Experimental setup. Noise from a tunnel
junction (green double box) biased at DC with a current $I_{N}$ couples
through capacitors $C_{1}=230\,$pF and $C_{2}=345\,$pF to a Josephson
junction detector (JJ, in orange), which is current-biased on its
supercurrent branch. The voltage $V_{J}$ across the junction monitors the
switching to the dissipative state. Capacitor $C_{J}=12.5\,$pF lowers the JJ
plasma frequency to $\protect\omega _{p0}/2\protect\pi \simeq 1.5$ GHz.
Capacitors $C_{N}=190\,$pF and $C_{g}=140\,$pF shunt the external impedance
at $\protect\omega _{p0}$, so that the impedance across the JJ is determined
only by on-chip elements. Resistors $R_{1}=R_{3}=215\,\Omega $ and $%
R_{2}=515\,\Omega $ were fabricated with thin Cr films. Bottom left: $IV$
characteristic of the tunnel junction, linear at this scale, with inverse
slope $R_{N}=22.9\,$k$\Omega $. Bottom right: $IV$ characteristic of the
detector JJ, with critical current $I_{0}=437\,$n$A$. We attribute a resonance near $V_{J}$ $\sim 120\,\mu $V
to a mode of the electromagnetic environment of the junction.} 
\label{setup}
\end{figure}

The experimental setup is shown schematically in Fig.\thinspace 1. As it is
well established that current noise through a tunnel junction is Poissonian (%
$S_{2}=e\left| I_{N}\right| $ and $S_{3}=e^{2}I_{N}$, with $e$ the electron
charge), we use such a device (green double box) as a benchmark noise
source. The detector JJ (orange crossed box) is coupled to it through
capacitors $C_{1}$ and $C_{2}.$ The finite frequency part $\delta I_{N}$ of
the current through the tunnel junction $I_{N}(t)$ flows through the
detector JJ, owing to the high-pass filter formed by $R_{3},$ $C_{1}$ and $%
C_{2}$ ($3\,$dB point at $5\,$MHz). The switching of the JJ current-biased
at $I_{J}$ is signaled by the appearance of a voltage $V_{J}$ across it. The 
low plasma frequency of 1.5\thinspace GHz guarantees $k_{B}T>\hbar
\omega _{p}/2\pi $ even at the lowest temperature of our experiment
(20\thinspace mK) \cite{BW}. In the relevant range of frequencies slightly
below $\omega _{p0}/2\pi $, numerical simulations of the actual circuit
indicate that the quality factor of the Josephson oscillations $Q$ is close
to $5,$ insuring an underdamped dynamics, and no effect of retrapping \cite%
{Melnikov,Timofeev} as long as $s\gg 4/\pi Q\simeq 0.25.$

The sample was fabricated on a thermally oxidized high resistivity (10$^{3}$
to 10$^{4}\,\Omega \,\mathrm{cm}$) Si wafer. All on-chip resistors are $10\,$%
nm-thick Cr layers, with $215\,\Omega /\Box $ sheet resistance at
4\thinspace K, placed between mm-size pads. Capacitors were obtained from
parallel aluminum films separated by 29\thinspace nm-thick sputtered silicon
nitride as an insulator \cite{notesurHuard}. The tunnel junction and the
detector JJ were fabricated at the same time by shadow evaporation of
20\thinspace nm and 80\thinspace nm-thick aluminum films. Their
current-voltage characteristics are shown in Fig.\thinspace 1. The tunnel junction has an area of $0.09\,%
{\mu}%
$m$^{2}$ and a tunnel resistance $R_{N}=22.9\,$k$\Omega .$ It was biased at
voltages larger than twice the superconducting gap $2\Delta /e=0.4\,$mV
(which corresponds to $I_{N}=0.02\,$%
$\mu$%
A), so that it behaves as a normal metal junction, with Poissonian noise.
The detector JJ, with area $1\,%
{\mu}%
$m$^{2},$ has a supercurrent $I_{0}=0.437\,$%
$\mu$%
A. It was biased in series with a resistor $R_{1}=215\,\Omega $ through a
50\thinspace $\Omega $ coaxial line equipped with attenuators. When
switching occurs at a supercurrent $I_{\text{sw}},$ the voltage across the
junction jumps to $\left( R_{1}+50\,\Omega \right) I_{\text{sw}}<2\Delta /e,$
so that the current through it drops to zero and no quasiparticles are
generated. Moreover, gold electrodes in good contact with the Al
films were fabricated a few 
$\mu$%
m away from the junctions in order to act as traps for spurious
quasiparticles \cite{traps} that could be excited by the high frequency
noise. Apart from the Cr resistors and the Au traps, all conductors on the
chip are superconducting aluminum films.

The sample was thermally anchored to the mixing chamber of a dilution
refrigerator. The tunnel junction was biased by a floating voltage supply through two $1.5\thinspace $M$\Omega $ resistors. The on-chip capacitance $%
C_{N}=190\,$pF on the bias line is large enough to maintain the voltage
across the tunnel junction at $V_{N}=R_{N}I_{N}$ for all relevant frequencies%
$.$ Escape rates of the JJ were measured using $2\times 10^{5}$ current
pulses of duration $\tau =0.53\,{\mu }s$ with alternatively positive ($%
+I_{J} $) and negative ($-I_{J}$) amplitude, separated by 9\thinspace 
$\mu$%
s. They were fed through a non-polarized capacitor $C_{\text{in}}=200\,$%
$\mu$%
F placed at room temperature, which prevents DC thermoelectric currents from
unbalancing the pulses. The switching rates $\Gamma _{+}$ and $\Gamma _{-}$
for the two signs of $I_{J}$ were deduced from the switching probability $%
P=1-e^{-\Gamma \tau }$ measured as the fraction of the current pulses which
led to a voltage pulse. 
\begin{figure}[tbp]
\includegraphics[width=3.2in]{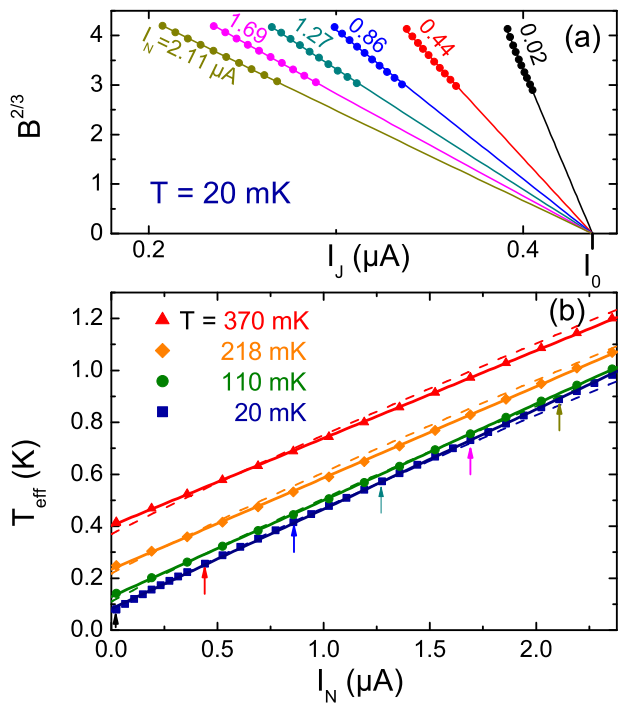}
\caption{(color online) (a) Dependence of $B^{2/3}=\left( -\log (\Gamma
/A)\right) ^{2/3}$ on the current $I_{J}$ through the detector JJ, for data
taken at $T=20\,$mK and currents through the tunnel junction $I_{N}=0.02$ to
2.11$\,{\protect\mu }$A, by steps of 0.42 ${\protect\mu }$A. Linear
dependence is a signature of the thermal activation regime. (b) Effective
temperature extracted from the slope of datasets as in (a), as a function $%
I_{N}$, for various temperatures $T$. Arrows indicate the datapoints
corresponding to the plots in (a). Solid lines are linear interpolations.\
Dashed line are the predictions from full theory, taking into account the
frequency dependence of the admittance $Y(\protect\omega )$ across the JJ
and of the transfer function $\protect\alpha (\protect\omega )$ from noise
source to JJ detector.}
\label{Tesc}
\end{figure}
\begin{figure}[tbp]
\includegraphics[width=3.3in]{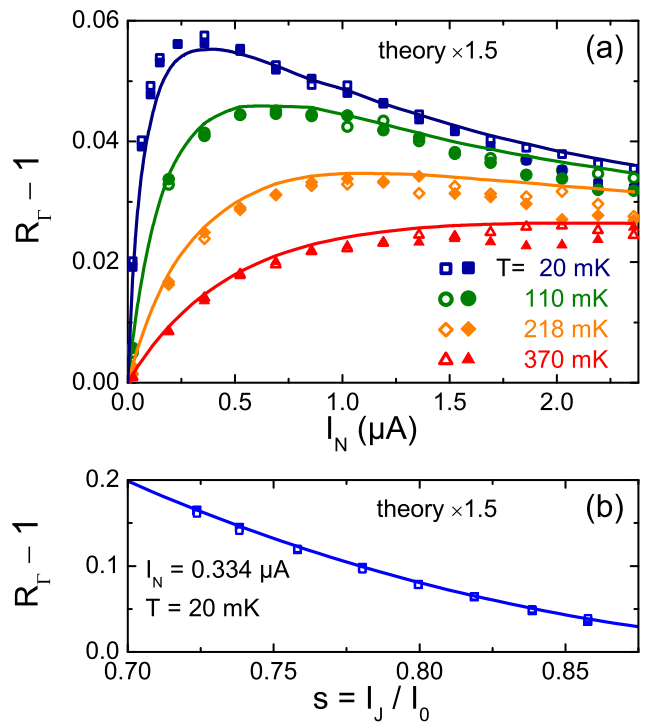}
\caption{(color online) (a) Rate asymmetry $R_{\Gamma }-1$ as a function of
current in noise source $I_{N}$ for various temperatures $T$. Filled (open)
symbols correspond to measurements with positive (negative) current through
the detector JJ. Solid lines are theoretical predictions scaled by a factor
1.5. (b) Dependence of $R_{\Gamma }-1$ on $s=I_{J}/I_{0}$, at $I_{N}=0.334\,{%
\protect\mu }A$, obtained with measurement pulses of various lengths. Solid
line is theoretical prediction scaled by 1.5.}
\label{Rgamma}
\end{figure}

We first demonstrate that the switching of the detector junction is well
described by the model of thermal activation whatever the current in the
noise source. Figure\thinspace 2(a) shows, for various currents $I_{N}>0,$
the $s-$dependence of $B^{2/3}=\left[ -\ln \left( \Gamma /A\right) \right]
^{2/3}.$ Data fall on straight lines that extrapolate to $0$ for $%
I_{J}=I_{0} $, as expected from Eq.$\,$(\ref{B2}). This allows us to extract
an effective temperature $T_{\mathsf{eff}},$ whose dependence on $I_{N}$ is
shown in Fig.\thinspace 2(b), with data taken at four different base
temperatures $T.$ We do find a linear dependance with correct extrapolations
at $I_{N}=0$ (values slightly above $T$ are attributed to imperfect
filtering), as expected from Eq.\thinspace (\ref{Teff}) with $S_{2}=eI$.
Understanding the slope quantitatively requires an accurate model of the
actual circuit at microwave frequencies: the RCSJ model assumes that the JJ
is simply connected to a capacitor, a resistance $R_{\parallel }$ and a
current source, in parallel. In the limit $Q\gg 1,$ $R_{\parallel },$ which
describes friction, has to be replaced with $R_{\parallel }(\omega
_{p})\equiv 1/\func{Re}\left( Y\left( \omega _{p}\right) \right) ,$ with $%
Y\left( \omega \right) $ the total admittance of the circuit across the JJ.
Microwave simulations indicate that $R_{\parallel }(\omega _{p})$ varies
almost linearly from 63$\,\,\Omega $ at 1\thinspace GHz to 36$\,\,\Omega $
at 1.5\thinspace GHz, and that a current $I_{N}(\omega )$ through the tunnel junction
leads to a current $\alpha (\omega )I_{N}(\omega )$ through the detector JJ$%
, $ with a transfer function $\alpha (\omega )$ varying from 1.1 at
1\thinspace GHz to 1.27 at 1.5\thinspace GHz. Since escape is determined
essentially by the noise at $\omega _{p},$ we replace $S_{2}$ by $\alpha
^{2}(\omega _{p})eI_{N}$. Altogether, the prediction $T_{\mathsf{eff}}\simeq
T+\alpha ^{2}(\omega _{p})R_{\parallel }(\omega _{p})eI_{N}/2k_{B}$ is in
agreement with the data (see dashed lines in Fig.\thinspace 2; to fit the
20\thinspace mK-data, we used $T=72\,$mK), apart from the slight change in
slope when varying $T$ which could be attributed to variations in the
kinetic inductance of the superconducting electrodes.

We now discuss the effect of noise asymmetry. The $B^{2/3}$ plots for
opposite signs of the current through the noise source are undistinguishable
within the symbol size, demonstrating that the effect of the second cumulant 
$S_{2}$ is dominant. In the limit $eV_{N}\gg k_{B}T$, theory predicts that
the effect of $S_{3}$ is to shift the curves by $\Delta I_{J}\sim
0.6B(\omega _{p0}/Q)(S_{3}/S_{2})=0.6Be/R_{\parallel }C_{J}\sim 0.2\%\times
I_{0}$, which is difficult to measure reliably \cite{notesurP}. In our
experiment, we measured directly the asymmetry ratio $R_{\Gamma }$ defined
by Eq.\thinspace (\ref{Rg}), which varies by several \% (see Fig.\thinspace
3). We first set the amplitude $I_{J}$ of the current pulses at a value
corresponding to a switching probability $P\sim 0.6,$ for which the
statistical precision on the rates is good \cite{correctionSQRT2}. We then
measured $100$ times $\Gamma _{+}$ and $\Gamma _{-}$, with alternatively $%
+I_{N}$ and $-I_{N}$ through the noise source. This allows for two
independent measurements of $R_{\Gamma }:$ $R_{\Gamma }^{+}=\Gamma
_{+}\left( -I_{N}\right) /\Gamma _{+}\left( +I_{N}\right) $ and $R_{\Gamma
}^{-}=\Gamma _{-}\left( +I_{N}\right) /\Gamma _{-}\left( -I_{N}\right) $ %
\cite{sign}$.$ In Fig.\thinspace 3(a), we show with full and open symbols
the corresponding measurements. The rate ratio $R_{\Gamma }$ differs from $1,
$ a signature of asymmetric noise, as soon as $I_{N}\neq 0$. The statistical
uncertainty on $R_{\Gamma }$ is smaller than the symbols. Small differences
between $R_{\Gamma }^{+}$ and $R_{\Gamma }^{-}$, in particular around $%
I_{N}=2\,$%
$\mu$%
A, are not understood. As for the comparison with theory, a difficulty
arises because of the frequency dependence of the transfer function $\alpha
(\omega )$, which results in a colored third cumulant at the detector $%
S_{3}(\omega _{1},\omega _{2})=\alpha (\omega _{1})\alpha (\omega
_{2}-\omega _{1})\alpha (-\omega _{2})e^{2}I_{N}.$ In the following, and in
the absence of indication as to which frequencies are important, we compare
however with the only existing theory, which assumes white noise ($%
S_{3}=e^{2}I_{N}$).\ The corresponding predictions, Eqs.\thinspace (\ref{B3},%
\ref{Rg}) with $j(s)\simeq 0.81\left( 1-s\right) ^{2.14}$\cite{notegamma},
are shown as solid lines in Fig.\thinspace 3(a), scaled by an arbitrary
factor 1.5. $R_{\Gamma }$ exhibits a maximum as a function of $I_{N}$ due to
the opposite variations of $S_{3}$ and $T_{\text{eff}}$ with $I_{N}.$ For $%
T_{\mathsf{eff}},$ we used interpolations between the measured values shown
in Fig.\thinspace 2. When scaled up by 1.5, which might be due to frequency
dependent transmission ($\alpha (\omega )$), theory accounts well for the
experimental data. Feedback corrections due to the detector, described in %
\cite{Sukhorukov}, are neglected since $R_{\parallel }/R_{N}\ll 1$ \cite%
{Grabert}. Note that there is no feedback associated to the series
resistance $R_{3}$ like in Ref.\thinspace \cite{Reulet} because the current
noise associated to $R_{3}$ does not flow through the noise source, but
through the detector JJ. In Fig.\thinspace 3(b), we also compare with theory
the $s-$dependence of $R_{\Gamma }$. In order to perform this measurement,
we used pulses of various durations (0.53 to 21\thinspace 
$\mu$%
s), which allows to obtain the switching rates at different values of $s.$
For the longest pulses, the rate asymmetry is as large as 16\%. Here also, theory scaled by 1.5 accounts precisely for the
data.

Qualitative agreement between experiment and theory gives confidence for the
use of the JJ as a measuring device for $S_{3},$ even if the application to
a wider range of systems requires some theory for colored noise. A
limitation concerns situations with strong non-linearities in the voltage
dependence of the cumulants, where feedback effects could become sizeable %
\cite{gap,Urban}. For quantitative measurements of $S_{3}$ on other systems,
it is not only important to tune the plasma frequency of the junction in the
GHz range as done in this work, but also to improve the microwave design, in
particular with more compact electrodes, so as to avoid frequency dependent
factors in the analysis. Proposals to access the full counting statistics
with a JJ embedded in more complex circuits \cite{Tobiska} remain to be
investigated.

\begin{acknowledgments}
We acknowledge technical support from Pascal Senat and Pief Orfila, and
discussions with B. Huard, H. Grabert, B.\ Reulet, J. Ankerhold, and within
the Quantronics group. Work supported by ANR contracts Electromeso and
Chenanom, and Region Ile-de-France for the nanofabrication facility at SPEC.
N.O.B. acknowledges support by NSF grant DMR-0705213.
\end{acknowledgments}


\begin{thebibliography}{99}
\bibitem{Levitov} L. Levitov and G. Lesovik, JETP Lett. \textbf{55}, 555
(1992).

\bibitem{Nazarov} Y. Nazarov, Ann. Phys. (Leipzig) \textbf{16}, 720 (2007).

\bibitem{Reulet} B. Reulet, J. Senzier and D.E. Prober, Phys. Rev. Lett. 
\textbf{91}, 196601 (2003).

\bibitem{Bomse} Yu. Bomze \textit{et al.}, Phys. Rev. Lett. \textbf{95},
176601 (2005).

\bibitem{Gershon} G. Gershon \textit{et al.}, Phys. Rev. Lett. 101, 016803
(2008).

\bibitem{Tobiska} J. Tobiska and Yu.V. Nazarov, Phys. Rev. Lett. \textbf{93}%
, 106801 (2004).

\bibitem{Timofeev} A.V. Timofeev \textit{et al.}, Phys. Rev. Lett. \textbf{98%
}, 207001 (2007).

\bibitem{Peltonen} J.T. Peltonen \textit{et al.}, Physica E \textbf{40}, 111
(2007).

\bibitem{Huard} B. Huard \textit{et al.}, Ann. Phys. (Leipzig) \textbf{16},
736 (2007).

\bibitem{Melnikov} V.I. Mel'nikov, Phys. Rep. \textbf{209}, 1 (1991).

\bibitem{Comparison} In the regime of thermal activation, the predictions of
this adiabatic model disagree by a factor 6 with those of the complete
theories of Ref.\thinspace \cite{Ankerhold,Sukhorukov,Grabert}.

\bibitem{Ankerhold} J.\ Ankerhold, Phys. Rev. Lett. \textbf{98}, 036601
(2007); \textit{ibid.}, \textbf{99}, 139901 (2007).

\bibitem{Sukhorukov} E.V. Sukhorukov and A.N. Jordan, Phys. Rev. Lett. 
\textbf{98}, 136803 (2007).

\bibitem{Grabert} H. Grabert, Phys. Rev. B \textbf{77}, 205315 (2008). We
corrected for a minus sign missing in Eqs.\thinspace (78,92).

\bibitem{Barone} A. Barone and G. Paterno, \textit{Physics and Applications
of the Josephson Effect} (Wiley, New York, 1982).

\bibitem{bookAnkerhold} For a review, see J. Ankerhold, \textit{Quantum
Tunneling in Complex Systems }(Springer, Berlin, 2007).

\bibitem{BW} The asymmetry signal increases with the detection bandwidth
determined by $\omega _{p}.$ However, it becomes more difficult to achieve a
frequency-independent coupling.

\bibitem{notesurHuard} The results of the experiment described in
Ref.\thinspace \cite{Huard} were presumably dominated by a leakage in
capacitors.

\bibitem{traps} P. Joyez \textit{et al.}, Phys. Rev. Lett. \textbf{72}, 2458
(1994).

\bibitem{notesurP} In Ref.\thinspace \cite{Timofeev}, the effect of $S_{3}$
was detected as the shift $\Delta I_{J}$. According to theory and using the
parameters given by the authors, the expected shift would have been much
larger than observed, had the thermal activation regime been achieved.

\bibitem{correctionSQRT2} The expression for the uncertainty on $R_{\Gamma }$
in Ref.\thinspace \cite{Huard} should be divided by $\sqrt{2}.$

\bibitem{sign} The signs in front of $I_{N}$ account for the fact that in
our setup, the noise current \textit{subtracts} from the bias current.

\bibitem{notegamma} This analytical expression interpolates between the
results of numerical calculations of $j(s)$ performed for $Q\sim $5 and in
the working interval $0.5 < s < 0.9,$ along the lines of Ref.\thinspace \cite%
{Grabert} (H. Grabert, private communication).

\bibitem{gap} In measurements with the noise source biased near $V=2\Delta
/e,$ where the $IV$ characteristic is very non-linear, we found strong
asymmetries in the switching rates.

\bibitem{Urban} D.F.\ Urban and H. Grabert, arXiv:0810.2938.

\end{thebibliography}
\end{document}